\documentclass[final,5p,times,twocolumn]{elsarticle}

\usepackage{graphicx}

\usepackage{amssymb}

\biboptions{sort&compress}

\begin{document}

\begin{frontmatter}


\title{Emission of forward neutrons by 158A~GeV indium nuclei
in collisions with Al, Cu, Sn and Pb}

\author[inr]{E.V.~Karpechev}
\author[inr]{I.A.~Pshenichnov\corref{cor1}}
\ead{pshenich@inr.ru}
\author[inr]{T.L.~Karavicheva}
\author[inr]{A.B.~Kurepin}
\author[inr]{M.B.~Golubeva}
\author[inr]{F.F.~Guber}
\author[inr]{A.I.~Maevskaya}
\author[inr]{A.I.~Reshetin}
\author[inr]{V.V.~Tiflov\fnref{Deceased}}
\author[inr]{N.S.~Topilskaya}

\author[DiSTA]{P.~Cortese}
\author[DiSTA]{G.~Dellacasa\fnref{Deceased}}

\author[unito]{R.~Arnaldi}
\author[unito]{N.~De~Marco}
\author[unito]{A.~Ferretti}
\author[unito]{M.~Gallio}
\author[unito]{A.~Musso}
\author[unito]{C.~Oppedisano}
\author[unito]{A.~Piccotti\fnref{Deceased}}
\author[unito]{E.~Scomparin}
\author[unito]{E.~Vercellin}

\author[infn]{C.~Cical\`{o}}
\author[infn]{G.~Puddu}
\author[infn]{E.~Siddi}

\author[CERN,ncbj]{P.~Szymanski}
\author[CERN]{I.~Efthymiopoulos}

\cortext[cor1]{Corresponding author}
\fntext[Deceased]{Deceased}

\address[inr]{Institute for Nuclear Research, Russian Academy of Sciences,
                 117312 Moscow, Russia}

\address[DiSTA]{Dipartimento di Scienze e Tecnologie Avanzate and INFN, Corso
Borsalino 54, 15100 Alessandria, Italy}

\address[unito]{ Universit\`{a} di Torino and INFN, Via Pietro Giuria 1,
   10125 Torino, Italy}

\address[infn]{ Universit\`{a}  di Cagliari and INFN, Casella Postale 170,
      09042 Monserrato (Cagliari), Italy}

\address[CERN]{CERN, EP Division, CH-1211,Geneva 23, Switzerland}

\address[ncbj]{Soltan Institute for Nuclear Studies, Warsaw, Poland}


\begin{abstract}
The cross sections of forward emission of one, two and three neutrons
by 158A~GeV $^{115}$In nuclei in collisions with Al, Cu, Sn and Pb targets are reported.
The measurements were performed in the framework of the ALICE-LUMI experiment at
the SPS facility at CERN. Various corrections accounting for the absorption of
beam nuclei and produced neutrons in target material and surrounding air were introduced.
The corrected cross section data are compared with the predictions of the RELDIS model
for electromagnetic fragmentation of $^{115}$In  in ultraperipheral collisions,
as well as with the results of the abrasion-ablation model for neutron emission in hadronic
interactions. The measured neutron emission cross sections well agree with
the RELDIS results, with the exception of In-Al collisions where the
measured cross sections are larger compared to RELDIS.
This is attributed to a relatively large contribution of hadronic fragmentation of In
on Al target with respect to electromagnetic fragmentation, in contrast to
similar measurements performed earlier with 30A~GeV $^{208}$Pb colliding with Al.
\end{abstract}


\end{frontmatter}



\section{Introduction}

The researches studying nucleus-nucleus collisions at the Super Proton Synchrotron (SPS) and
at the Large Hadron Collider (LHC) at CERN are mostly focused on the studies of hot and dense
nuclear matter, which is created in collisions of ultrarelativistic nuclei. This includes the search of
signals of the phase transition between the hadronic matter and the quark-gluon phase which can be
produced in collisions with small impact parameters. Such central collisions are characterized
by a large overlap of the volumes of colliding nuclei, their complete disintegration and very high
multiplicities of produced secondary particles.

The studies of collisions of ultrarelativistic nuclei remain incomplete without exploring the domain
of large impact parameters. There is another phenomenon which deserves attention there, namely,
ultraperipheral interactions of nuclei which occur without any overlap of
their volumes.  The disintegration of nuclei in collisions with impact parameters $b\geq R_1+R_2$,
where $R_1$ and $R_2$ are the nuclear radii, can only be explained by the long-range electromagnetic forces.

The impact of the Coulomb field of a fast-moving nucleus with the charge $Z$ on its collision partner
can be estimated as following. Providing that in the rest frame of the collision partner
the moving nucleus has the Lorentz factor $\gamma$, the Lorentz-boosted Coulomb potential
at the moment of the closest approach can be
estimated as $V_{c}=\alpha \gamma Z/b$, where $\alpha$ is the fine structure constant.
Therefore, for collisions of medium-weight nuclei ($Z\sim 50$, $b\sim 10$~fm) with beam
energies available at the CERN SPS the relation $\alpha\gamma\sim 1$ holds and $V_{c}\sim 1$~GeV.
The estimated $V_{c}$ essentially exceeds the effective nuclear potential
$\sim 50$ MeV, which confines nucleons inside nuclei.
This explains the emission of single nucleons from nuclei and their fragmentation
as a result of electromagnetic interactions in ultraperipheral collisions.
This phenomenon is well-known as electromagnetic dissociation (EMD) of
nuclei~\cite{BertBaur,Baur:2001jj,PR2008}.
The behavior of nuclear matter under the impact of strong electromagnetic fields
can be studied in ultraperipheral collisions of ultrarelativistic nuclei.

The Lorentz contraction of electromagnetic fields of nuclei in ultraperipheral nucleus-nucleus
collisions at the LHC becomes tremendous. As demonstrated by recent
measurements~\cite{Oppedisano:2011rx,AbelevALICE2012}, the total cross section of neutron emission
from at least one of the colliding lead nuclei approaches 187~b at 1.38A~TeV$+$1.38A~TeV collision energy.
In this case the nucleus which impacts another nucleus via the action of its electromagnetic
field is characterized by a large Lorentz factor $\gamma\sim 4.3\times 10^6$ which
defines the scale of contraction.

As demonstrated in several publications, the EMD process plays a certain role at the LHC.
On the one hand a large EMD cross section imposes restrictions on the beam
lifetime at the LHC~\cite{Baltz}.
Nuclear fragments produced in EMD events can lead to local heating of the LHC
construction elements~\cite{Klein,Bruce}. On the other hand, the collider luminosity can be
monitored~\cite{Baltz2} by counting mutual EMD events characterized by
the simultaneous emission of forward neutrons by each of the nuclei in a single ultraperipheral event.
As suggested~\cite{Pshenichnov:2001qd,PR2008},
the rate $R_m^{EMD}$ of mutual EMD events with the emission of either one or two neutrons
in the directions of each beam should be measured at the LHC. This makes
possible to measure the collider luminosity
$L=R_m^{EMD}/\sigma_m^{EMD}$ providing that the mutual EMD cross section $\sigma_m^{EMD}$
is calculated with sufficient accuracy~\cite{Pshenichnov:2001qd,PR2008,Pshenichnov:2011}.

The single and mutual EMD neutron emission cross sections can be calculated with
the RELDIS Monte Carlo model~\cite{Pshenichnov,Pshenichnov2,Pshenichnov:2001qd,Pshenichnov:2011}, which is
based on the Weiz\"acker-Williams method~\cite{BertBaur}.
In order to ensure the accuracy of this model for Pb-Pb collisions at the LHC, its results were
validated at lower collision energies. In particular, we have measured the cross sections
of forward neutron emission from 30A~GeV lead nuclei in collisions with various target
nuclei~\cite{Golubeva}.  As proved by these measurements~\cite{Golubeva}, the cross sections of forward
emission of one and two neutrons are well described by the RELDIS model.
Recently the single and mutual EMD cross sections of neutron emission were measured at the LHC for Pb-Pb collisions
at 1.38A~TeV$+$1.38A~TeV~\cite{Oppedisano:2011rx,AbelevALICE2012}, and a very good agreement with the RELDIS
results was demonstrated.

In the future the research programs at the CERN-SPS and LHC can be extended to study collisions of
medium-weight nuclei. It is the purpose of the present work to study the electromagnetic
dissociation of indium  nuclei in ultrarelativistic collisions with
aluminum, copper, tin and lead target nuclei. The cross sections of forward emission of one,
two and three neutrons by 158A~GeV $^{115}$In nuclei are measured and compared to
the corresponding results of the RELDIS model. The fragmentation of 158A~GeV indium nuclei
in collisions with Si, Ge, Sn, W and Pb was studied in Ref.~\cite{Uggerhoj}  by
measuring the cross sections of production of secondary fragments with specific charge, also known as
the charge-changing cross sections. Since neither fragment masses, nor neutrons emitted from
indium nuclei were identified in Ref.~\cite{Uggerhoj}, the measured cross sections represented
the sum of hadronic fragmentation cross sections due to direct overlap of nuclei and fragmentation
cross sections in ultraperipheral collisions.
In contrast to Ref.~\cite{Uggerhoj}, mainly the contribution of electromagnetic dissociation was
identified in our experiment due to the detection of forward neutrons from
indium nuclei, as explained below. In this sense our data complement the data collected in 
Ref.~\cite{Uggerhoj} and provide specific information on electromagnetic interactions of 
ultrarelativistic $^{115}$In nuclei.

\section{Neutron emission from indium nuclei in collisions with target nuclei}\label{photonuclear}

We consider now the neutron emission from indium nuclei resulting from electromagnetic and
hadronic interactions of these nuclei with various targets. Once theoretical models describing
such processes are at hand, the relations between the cross sections of such  processes
can be estimated along with energy and angular distributions of produced neutrons. This helps in
estimating the parameters of our experimental setup which is designed to detect such neutrons.

\subsection{Neutrons from electromagnetic dissociation of indium nuclei}

The Weizs\"acker-Williams method of virtual photons~\cite{BertBaur,Baur:2001jj,PR2008} is widely used
to describe ultraperipheral interactions of nuclei. According to this method electromagnetic
interactions of nuclei proceed by emission and absorption of virtual photons by these nuclei.
The range of equivalent photon energies from the nucleon emission threshold
$E_{min}$ to $E_{max}\sim\gamma/(R_1+R_2)$~\cite{Pshenichnov:2011} should be taken into account
in order to describe the electromagnetic dissociation of nuclei.  According to
the threshold of $(\gamma ,n)$ reaction on $^{115}$In, $E_{min}=9.26$~MeV was adopted in our calculations.
The maximum photon energy $E_{max}$ depends on the sizes of colliding nuclei:
$E_{max}=3.32$ and 2.36~GeV for ultraperipheral interactions of 158A~GeV $^{115}$In
with Al and Pb nuclei, respectively. The equivalent photon spectrum calculated with various
assumptions is given in Refs.~\cite{BertBaur,Baur:2001jj,PR2008}.

The RELDIS model uses the Monte Carlo method to simulate EMD events. This model
is described in Refs.~\cite{Pshenichnov,Pshenichnov2,Pshenichnov:2001qd,Pshenichnov:2011}.
The absorption and emission of one and two photons in a single ultraperipheral collision is taken into
account, and the hadronic production of secondary particles in each collision is
simulated by the Monte Carlo method. Depending on the energy $E$ of a photon which is
absorbed by a nucleus, various processes take place, namely, the excitation of a giant resonance
in this nucleus, the absorption of the photon by a pair of bound nucleons or the
production of hadrons on a bound nucleon.

It is expected that the excitation and decay of the giant dipole resonance (GDR) at
$9\leq E \leq 35$~MeV provides the main contribution to electromagnetic dissociation
of indium nuclei. Therefore, below we briefly review the modeling of the GDR excitation
and decay within the RELDIS model. This process usually leads to the emission of one or two
neutrons from In nucleus. The fission of In is not probable, and the cross section of
emission of protons following the absorption of low-energy photons is also small due to
a high Coulomb barrier in this medium-weight nucleus. The total probability of photon absorption
is calculated as an integral of the product of the equivalent photon spectrum and
the total photoabsorption cross section~\cite{Pshenichnov:2001qd,Pshenichnov:2011}.

\begin{figure}[tb]
\includegraphics[width=1.0\columnwidth]{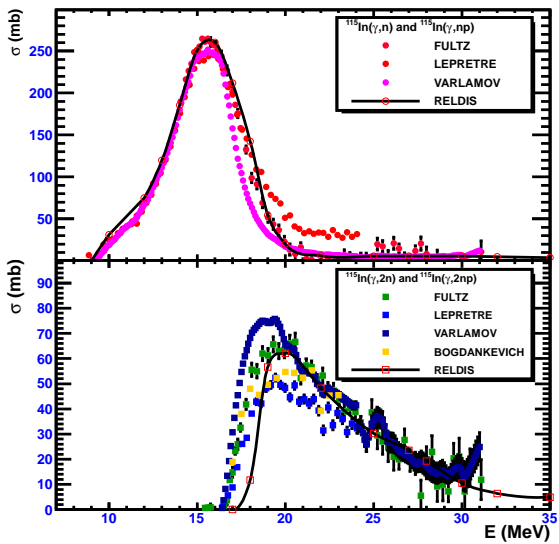}
\caption{Cross sections of emission of one (top panel) and two neutrons (bottom panel)
by $^{115}$In as a result of absorption of a photon with energy $E$.
Experimental data~\cite{Fultz:1969,Lepretre:1974,Bogdankevich:1962} and
evaluated data~\cite{Varlamov:2010} all extracted from the EXFOR database~\cite{EXFOR} are shown by various
symbols explained on the legends of the corresponding panels. Results of the RELDIS code are shown
by open symbols connected by a solid line.
}
\label{fig1:gIn}
\end{figure}

It is assumed that the energy of a photon of $9\leq E \leq 35$~MeV is completely transformed into the excitation
energy of the nucleus which absorbs this photon. The subsequent evolution of an excited nucleus
is described by the Statistical Multifragmentation Model (SMM)~\cite{JPB}, which is widely used to describe decays
of excited nuclear systems. As predicted by the RELDIS model, the mean excitation energy
of a medium-weight or heavy excited system, e.g., In or Pb, after the absorption of a photon
is below 2 MeV/nucleon ~\cite{Pshenichnov:2001qd,Pshenichnov:2011}.
The SMM model predicts that the de-excitation of such systems proceeds via a sequential
evaporation of nucleons. In particular, there is a competition  between the emission of one
and two neutrons in the GDR energy region, see Fig.~\ref{fig1:gIn}.
As only neutrons were detected in experiments carried out in several
laboratories~\cite{Bogdankevich:1962,Fultz:1969,Lepretre:1974},
$(\gamma,n)$ and $(\gamma,2n)$ channels were not separated from $(\gamma,np)$ and $(\gamma,2np)$,
respectively. Therefore, their sum is presented in Fig.~\ref{fig1:gIn} at $E\leq 35$~MeV
and will be referred below as $(\gamma,n)$ and $(\gamma,2n)$, respectively, in order to simplify notations.

The most reliable evaluated nuclear data on $(\gamma,n)$ and $(\gamma,2n)$ cross sections on $^{115}$In
were obtained in Ref.~\cite{Varlamov:2010} based both
on theoretical results and measurements. The total photoabsorption cross section was estimated from
measurements~\cite{Bogdankevich:1962,Fultz:1969,Lepretre:1974}, while the pre-equilibrium model of
photonuclear reactions was used to disentangle the contributions of $(\gamma,n)$ and
$(\gamma,2n)$ processes.

In the RELDIS model the total photoabsorption cross section in the Giant Resonance region
($E\leq 40$~MeV) is calculated according to the approximations found in
Refs.~\cite{Berman-Fultz,Dietrich} for various medium-weight and heavy target nuclei, including In.
The relations between $(\gamma,n)$ and $(\gamma,2n)$
channels are calculated by RELDIS from the statistical evaporation model. As seen
from Fig.~\ref{fig1:gIn}, the experimental data on $(\gamma,n)$ reaction on In collected
at the Livermore National Laboratory~\cite{Fultz:1969} are described by RELDIS in general.
However, the threshold of $(\gamma,2n)$ reaction is overestimated. Nevertheless, the RELDIS results on
$(\gamma,2n)$ reaction remain closer to Livermore data~\cite{Fultz:1969}
compared to other measurements~\cite{Lepretre:1974,Bogdankevich:1962}.
At the same time on the right-hand side of the GDR resonance
the RELDIS results on $(\gamma,n)$ reaction are higher compared to the corresponding
evaluated cross section data~\cite{Varlamov:2010}.
An opposite trend is seen for the evaluated $(\gamma,2n)$~\cite{Varlamov:2010} cross section.
It is higher compared to the same cross section calculated by RELDIS.

Neutrons emitted by excited ultrarelativistic nuclei created in ultraperipheral collisions have very 
forward angular distributions in the lab frame. This is because such neutrons have few MeV kinetic energy 
and they are evaporated isotropically in the rest frame of the excited nucleus. 
After the Lorentz boost from this system to the laboratory system
such neutrons have velocities close to the velocity of the beam nuclei and they are characterized by 
small transverse momenta $P_t$ perpendicular to the beam direction. 
Distributions of $P_t$ for first and second neutrons emitted in electromagnetic dissociation of lead nuclei 
were calculated with RELDIS~\cite{Golubeva}. They have a maximum at $P_t\sim 0.02$--$0.04$~GeV/c
and a tail extending up to $P_t\sim 0.15$~GeV/c~\cite{Golubeva}. 
Similar distributions, also restricted by $P_t<0.15$~GeV/c, are calculated with RELDIS for 158A~GeV $^{115}$In nuclei.
They indicate that all EMD neutrons are emitted within $\pm 1$~mrad angle with respect to the beam direction. 
The location of the neutron calorimeter (1350~cm downstream from the target) and its 
transverse extensions ($\sim$3.5~cm at each direction perpendicular to the beam axis) ensure the detection of
all neutrons produced in EMD of 158A~GeV $^{115}$In nuclei, as a wider angle of $\pm 2.6$~mrad is covered by the
detector.

\subsection{Neutron emission in nuclear reactions induced by indium nuclei}

Hadronic interactions of nuclei are characterized by strong interactions between participating
nucleons in the overlapping parts of colliding nuclei.
The abrasion-ablation model provides a simplified description of such collisions, and its
modern version was already used to describe the fragmentation of
lead nuclei~\cite{Scheidenberger} with the same (158 GeV) beam energy per nucleon as of
indium beam in the present work.

Following the abrasion-ablation model nucleons from colliding nuclei are classified
into participants and spectators according to their role in the collision~\cite{Scheidenberger}.
Participant nucleons interact with nucleons from collision partner. Spectator nucleons
represent relatively cold spectator nuclear matter and do not interact with nucleons of
collision partner at the first abrasion stage of the collision. The number of participant nucleons
is calculated according to the Glauber theory of multiple scattering. Other details of the
abrasion-ablation model used in the present work are given elsewhere~\cite{Scheidenberger}.

In Ref.~\cite{Scheidenberger} various methods to calculate the excitation energy of
residual nuclei composed of spectator nucleons were presented. Hereafter such nuclei are called
prefragments. As shown~\cite{Scheidenberger}, the excitation energy of a prefragment can
be estimated as 13--26~MeV on average per each participant nucleon knocked out from
the initial nucleus. Excited prefragments decay at the
ablation stage of the collision and their decays in most cases are simulated as evaporation of
nucleons according to the above-mentioned statistical model SMM~\cite{JPB}.

Neutrons are produced on both stages of interaction, but their kinematic
characteristics are different. Nucleons are knocked out from initial nuclei as
a result of individual nucleon-nucleon collisions at the abrasion stage,
and their emission is accompanied by production of secondary hadrons. The average transverse momentum
of nucleons knocked out from nuclei is estimated as
$\langle P_t \rangle \sim 0.2-0.4$~GeV/c~\cite{Scheidenberger}, and it is essentially larger than the
average transverse momentum of neutrons produced in electromagnetic dissociation of nuclei.
However, neutrons emitted by excited prefragments at the ablation stage have a narrow
$P_t$-distribution, similar to the distribution of neutrons from electromagnetic dissociation.
Such ablation neutrons can be also registered by the calorimeter used in the present work.

Hadronic nucleus-nucleus collisions lead to emission of various numbers of neutrons in each event.
As expected, the average number of knocked out and emitted neutrons per event
depends on the collision centrality, i.e. on the impact parameter $b$. Since
the fragmentation events of indium nuclei which specifically lead to the emission of one, two and three
neutrons is the main subject of our study, peripheral (grazing) nucleus-nucleus collisions with
$b\sim R_1+R_2$ have to be considered. The number of participant nucleons in such peripheral collisions is
small and such nucleons are mostly located in a thin surface layer of each nucleus. This explains the fact
that the cross section of emission of a small number of nucleons amounts to a small fraction of the
total fragmentation cross section. In addition to the direct knock-out of nucleons at the abrasion stage
the evaporation of nucleons from excited residual nuclei takes place also at the ablation stage.
In this case the events with emission of only few neutrons at the ablation stage are
mostly classified as peripheral collisions characterized by relatively low excitation of
spectator matter.

Despite of the simplifications adopted in the above-described modeling of nucleus-nucleus collisions,
a satisfactory agreement of calculated and measured yields of nuclear
fragments can be expected. As shown in Ref.~\cite{Scheidenberger}, this approach
applied to peripheral collisions of 158A~GeV lead nuclei with C, Al, Cu, Sn and Au
successfully describes the yields of nuclear fragments which are close in mass and charge
to projectile nuclei. Such fragments are created after
emitting a few nucleons, and the nucleon multiplicity in such reactions is accordingly
described. Therefore, one can expect that the abrasion-ablation model is also applicable to
neutron emission by indium projectiles with the same beam energy per nucleon.

Calculated cross sections of emission of one, two and three neutrons, possibly
accompanied by other undetected particles, e.g. protons, are collected in Table~\ref{table1:theory}.
The total EMD and hadronic fragmentation cross sections are also given in this table.
Due to the presence of channels with emission of more than three neutrons in electromagnetic and
hadronic interactions, the sum of 1nX, 2nX and 3nX cross sections is less than the corresponding total
cross section. The sum of 1nX, 2nX and 3nX EMD channels amounts to 87\% of the total EMD cross section,
while the same channels of hadronic fragmentation provide only 6--8\% of the total
hadronic cross section.
\begin{table}[htb]
\begin{center}
\caption{Calculated cross sections of emission of one, two and three neutrons
and the total cross sections for electromagnetic dissociation and hadronic
fragmentation of 158A~GeV $^{115}$In on Al, Cu, Sn and Pb nuclei.
Results of the RELDIS and abrasion-ablation models are given for
electromagnetic and hadronic interactions of nuclei.}
\begin{tabular}{|c|c|c|c|}
\hline
Target  &  Fragmentation    &   EMD,   & hadronic          \\
nucleus & channel  &  RELDIS  & fragmentation,     \\
       &         &          & abrasion-ablation \\
       &         & (barns)  & (barns) \\
\hline
       &   1nX   &   0.40   &      0.06         \\
  Al   &   2nX   &   0.09   &      0.08         \\
       &   3nX   &   0.03   &      0.11         \\
       &  total  &   0.60   &      3.34         \\
\hline
       &   1nX   &   1.91   &      0.07         \\
  Cu   &   2nX   &   0.45   &      0.10         \\
       &   3nX   &   0.13   &      0.12         \\
       &  total  &   2.86   &      4.25         \\
\hline
       &   1nX   &   5.47   &      0.08         \\
  Sn   &   2nX   &   1.28   &      0.11         \\
       &   3nX   &   0.38   &      0.13         \\
       &  total  &   8.19   &      5.14         \\
\hline
       &   1nX   &   14.05  &      0.09         \\
  Pb   &   2nX   &    3.32  &      0.12        \\
       &   3nX   &    1.03  &      0.15         \\
       &  total  &   21.07  &      6.18         \\
\hline
\end{tabular}
\label{table1:theory}
\end{center}
\end{table}

\section{Experimental setup}

\begin{figure}[tb]
\includegraphics[width=1.0\columnwidth]{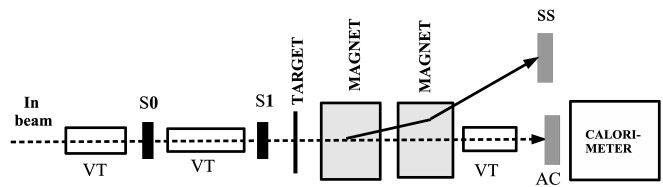}
\caption{Experimental setup to study neutron emission in electromagnetic fragmentation of
158A~GeV $^{115}$In. S0, S1, SS, AC - scintillator detectors, VT - vacuum tubes.
}
\label{fig2:Setup}
\end{figure}

Measurements of the cross sections of emission of forward neutrons in fragmentation
of 158A~GeV indium nuclei were performed at H8 beam line of the CERN-SPS accelerator
with a setup shown in Fig.~\ref{fig2:Setup}. The beam of $^{115}$In nuclei of
158A~GeV energy was focused on a target which can be moved into and out of the beam by means of
a mobile support.

Two plastic scintillator detectors, S0 and S1, were installed in front of the target. Both
detectors were made of polystyrene with addition of 4--5\% of PTP~POPOP. They were manufactured as thin plates
each 2~mm thick and of 2~cm$\times$2~cm size. Both detectors have demonstrated a very good energy resolution.
The main peak from 158A~GeV indium nuclei was clearly distinguished in the signals taken from both
detectors. There were additional tails related to a lower energy deposition in the scintillator plates
due to nuclear fragments produced in beam fragmentation. A small tail was also identified in the signal
obtained from the S0 detector due to the presence of nuclear fragments created in interactions of beam
nuclei with air and windows of vacuum tubes on their way to S0. Production of secondary particles in the beam line 
upstream of the S0 detector also contributed to this tail.

Charged particles were deflected beyond the acceptance of the neutron calorimeter by means of two
magnets of 4.4~T$\times$m each. These magnets were placed after the target and they deflected particles 
in the horizontal direction. Since noninteracting beam nuclei were deflected by 7~cm from the primary 
beam direction, it was possible to place there a plastic scintillator SS for their detection, as well 
as for the detection of $^{114}$In, $^{113}$In and $^{112}$In created in EMD.
These nuclei were deflected by the magnets by angles which were very close to the deflection angle of 
the primary beam. Since the SS detector had small dimensions of 2~cm$\times$2~cm thus covering a small range of
defection angles, a certain selectivity of SS to indium remnants created in EMD was achieved. 
Most of nuclear fragments created in hadronic fragmentation of $^{114}$In are expected to have charge-to-mass 
ratios which differ from the corresponding ratio of beam nuclei and therefore do not hit the SS detector. 

Neutrons from fragmentation of In nuclei were not deflected by magnets and therefore hit
the neutron calorimeter which was installed after the magnets directly on the axis of the primary beam.
The neutron calorimeter was placed on a platform with their surface aligned parallel to the beam axis.
An additional detector AC was installed downstream of the target in front of the neutron calorimeter.
AC was built as a scintillator detector, it was 2~mm thick with the transverse dimensions of
70~mm~$\times$~100~mm. As AC was fired by charged particles, it was used as a veto detector to
suppress the events in neutron calorimeter which were due to such particles.
The number of primary indium nuclei was defined by counts collected from the S1 detector, while the
SS detector was used to tag indium nuclei which did not fragment neither in air, nor in the target.

Four targets made of different materials (aluminum, cooper, tin and lead) were used. In addition
the fifth lead target was also used, but with increased thickness. It had the thickness about
twice as large as the first lead target. The corrections for the absorption of produced neutrons and
multiple interactions of indium nuclei in target material depend on the target thickness. However, providing that
such corrections are properly introduced, the cross sections of In-Pb interactions which are
calculated from the data collected with thin and thick targets are expected to be equal.
A possible difference of resulting cross sections can be considered as a systematic uncertainty of our
measurements. The parameters of the targets used in our experiment are listed in Table~\ref{materials}.
\begin{table}[htb]
\begin{center}
\caption{Atomic number $Z$, atomic mass $A$, material density $\rho$ and thickness $d$ of the
targets used in measurements with 158A~GeV $^{115}$In beam.}
\begin{tabular}{|c|c|c|c|c|c|}
\hline
&  Al & Cu & Sn & Pb thin & Pb thick \\
\hline
$\rho$ (g/cm$^3$) & 2.7 & 8.96 & 7.31 & 11.35 & 11.35 \\
\hline
$d$ (cm) & 1.4 & 0.45 & 0.34 & 0.135 & 0.294 \\
\hline
    $Z$ &   13 &  29   & 50   & 82 & 82 \\
\hline
    $A$ &   27 & 64   & 119  & 208 & 208 \\
\hline
\end{tabular}
\label{materials}
\end{center}
\end{table}

During a separate empty-target run only neutrons produced in the beam line collimators, scintillator
plates S0 and S1, and also in air were detected by the neutron calorimeter.
This provided us the estimation of the neutron background
due to nuclear reactions which take place beyond the target. Then this background was subtracted from the signal
obtained in other runs with installed targets.

There were several advancements of the described experimental setup with respect to our
previous study~\cite{Golubeva} of fragmentation of 30A~GeV lead nuclei in ultraperipheral collisions
with same target nuclei. Namely, the following improvements were achieved:
\begin{itemize}
\item  Vacuum tubes (VT) were placed before and after the S0 detector and also
between the magnets and neutron calorimeter. This helped us to reduce, respectively,
the numbers of beam nuclei and produced neutrons lost in their interactions with air.

\item The AC detector was
installed in front of the neutron calorimeter and was used as a veto detector.

\item Almost all neutrons from electromagnetic fragmentation of beam nuclei were covered by
the acceptance of neutron calorimeter. There was no need in introducing any corrections for a limited
acceptance in contrast to our previous measurements, see Ref.~\cite{Golubeva}.
\end{itemize}

\section{Response of the neutron calorimeter to pion and proton beams}

\begin{figure}[tb]
\includegraphics[width=1.0\columnwidth]{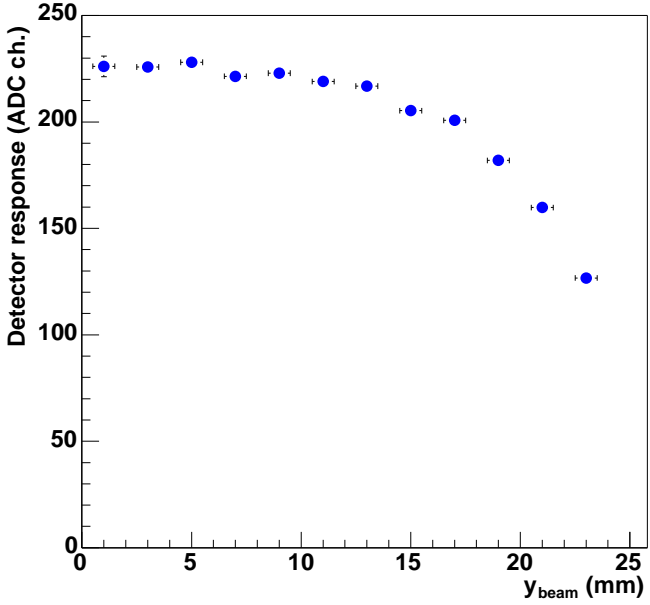}
\caption{Response of the calorimeter to 100~GeV pions as a function of
transverse shift of the beam axis $y$ with respect to the calorimeter axis.
}
\label{fig3:resp_y}
\end{figure}

\begin{figure}[tb]
\includegraphics[width=1.0\columnwidth]{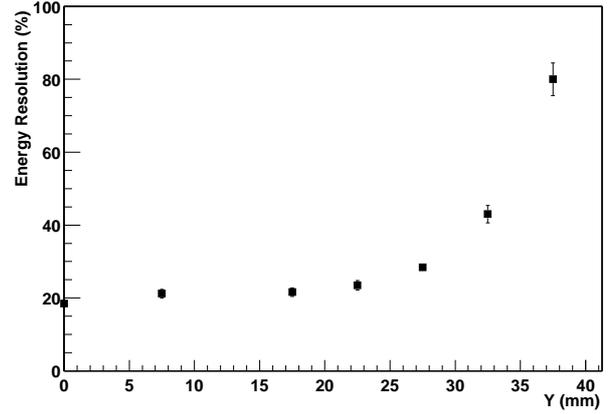}
\caption{Relative energy resolution of the calorimeter to 30~GeV pions as a function of
transverse shift $y$ with respect to the calorimeter axis.
}
\label{fig4:en_y}
\end{figure}

\begin{figure}[tb]
\includegraphics[width=1.0\columnwidth]{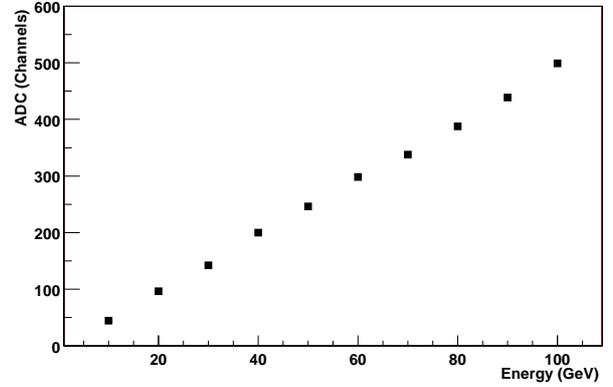}
\caption{Response of the calorimeter as a function of pion energy.
}
\label{fig5:en}
\end{figure}

\begin{figure}[tb]
\includegraphics[width=1.0\columnwidth]{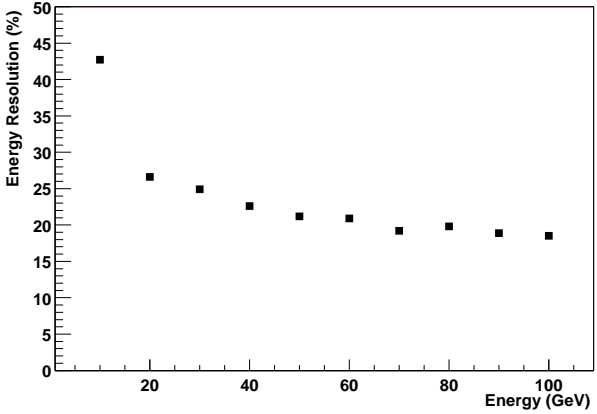}
\caption{Relative energy resolution of the calorimeter as a function of pion energy.
}
\label{fig6:en}
\end{figure}

\begin{figure}[tb]
\includegraphics[width=1.0\columnwidth]{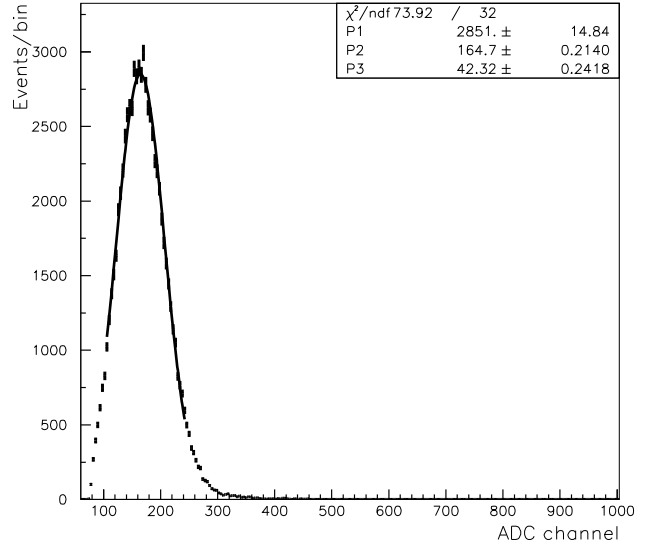}
\caption{Energy spectrum of the calorimeter for 30~GeV protons.
}
\label{fig7:en_prot}
\end{figure}

The neutron Zero Degree Calorimeter (ZDC) was a key part of the experimental setup designed for the
ALICE-LUMI experiment. Its performance was crucial for accurate measurements of neutron yields.
Therefore, before the main experiment with indium beam ZDC was tested with beams of other
particles of comparable energy.
The response functions of the neutron calorimeter to pions and protons are presented in
Figs.~\ref{fig3:resp_y}, ~\ref{fig4:en_y}, ~\ref{fig5:en}, ~\ref{fig6:en} and ~\ref{fig7:en_prot}.
The following design requirements are essential for effective registration of neutrons and reliable
determination of their multiplicity in each event:
\begin{itemize}
\item The response of neutron calorimeter remains constant over the transverse
section of this detector with its spot size covering the whole range of transverse momenta of
neutrons produced in EMD of beam nuclei. The detector response and relative energy resolution
as a function of the transverse shift $y$ of the beam with respect to the central
axis of the calorimeter are presented for 100 and 30 GeV pions,
in Figs.~\ref{fig3:resp_y} and ~\ref{fig4:en_y}, respectively.

\item The response of the neutron calorimeter is proportional to the projectile energy.
This makes possible to disentangle peaks corresponding to one, two and three neutrons.
The response of the calorimeter to pions as a function of pion energy is shown
in Fig.~\ref{fig5:en}.

\item The energy resolution of the calorimeter for neutrons emitted in the EMD process of beam nuclei
is sufficient for separation of events with different numbers of neutrons. This is proved by measurements
with pion beam. As shown in Fig.~\ref{fig6:en}, the energy resolution improves with the increase of pion energy.
As follows from the extrapolation of data shown in Fig.~\ref{fig6:en}
to 158~GeV neutrons from EMD, the relative energy resolution for this case will be better than 20\%.
\end{itemize}

The design of the neutron calorimeter was based on results of Monte Carlo simulations.
The calorimeter was assembled as a periodic structure consisting of 40 layers.
Each layer was made of three tungsten plates. Each tungsten plate was 2.5~mm thick and paved with
a 2~mm thick plastic scintillator plate, all of 7~cm$\times$10~cm dimensions.
As each layer was tilted by 45 degrees with respect to the beam axis in the vertical direction, 
the transverse cross section of the calorimeter was 7$\times$7.07~cm$^2$.
The collected light from all scintillator plates was re-emitted into 80 plastic fibers of 1~mm in diameter each
glued into the semicircular groves made on the sides of each scintillator plate. These fibers transported light 
through a fiber bundle to an XP2020 photomultiplier. The opposite ends of the optical fibers 
were covered by aluminum film in order to reduce the loss of light.

The X5 beam channel of the CERN SPS was used to calibrate the calorimeter by irradiating it
by pions with energy from 10 to 100~GeV and by 30~GeV protons. Two compact scintillator counters
of 2~mm$\times$2~mm size were used to identify the position of the beam on the forward surface
of the calorimeter. The distribution of signal collected from the calorimeter was approximated by
a Gaussian function with the amplitude, peak position and width defined by
fitting this distribution.

As shown in Fig.~\ref{fig5:en}, the response of the calorimeter to pions with energy from
10 to 100~GeV is proportional to pion energy.
The relative energy resolution $\sigma/\mu$ of $\sim$25\% for 30~GeV protons,
Fig.~\ref{fig7:en_prot}, was measured in a dedicated run which
provided the energy calibration for the calorimeter. During this run the magnets were switched off.

During another run with a 100~GeV pion beam
the response of the calorimeter was measured depending on the beam shift along the horizontal and
vertical directions.

\section{Measurements with $^{115}$In and data processing}

The main set of measurements of fragmentation of $^{115}$In was performed with a medium-intensity
beam of $5\times 10^5$ ions per spill. Three different triggers were used in measurements:
(1) a random coincidence trigger to estimate pedestal and noise; (2) a beam trigger only with S0;
and (3) a physics trigger for measurements of $^{115}$In fragmentation. The main trigger included
signals from S0, S1 and SS (S0$\times$S1$\times$SS) under the condition that the energy
registered by the neutron calorimeter exceeds 8~GeV. In addition, some measurements were performed
with S0$\times$SS trigger. The data collected in such measurements were not used for
the determination of the neutron emission cross sections, but rather were used to estimate the
corrections to raw data applied in processing data off-line.

The threshold settings of the S0, S1 and SS counters were estimated in runs without target.
In these runs the main signal in each of these detectors was due to beam nuclei and the thresholds were 
adjusted accordingly to avoid the rejection of beam particles and provide similar counting rates. 
Then the same settings were applied to the S0, S1 and SS counters during runs with target.

During each run the output from the neutron calorimeter was recorded into a separate file  
containing up to $10^6$ events. Each run had its own pedestal run to ascertain an accurate 
pedestal subtraction. The pedestal corrected data from all runs with a given target were combined 
into a single data set, after the ADC counts were converted into energies.
Each set was used afterwards to estimate neutron emission cross sections for the 
corresponding target.

It is expected that noninteracting beam nuclei and those which lost only few neutrons
after EMD produce similar Gaussian-shape spectra in the S0 and SS detectors.
Therefore, the parts of the spectra deviating from this Gaussian function can be attributed to 
the hadronic interactions of indium nuclei with the target, with air or with the windows of the 
vacuum tubes. Indeed, the energy per nucleon
of secondary nuclear fragments created in hadronic interaction can be quite different from the
beam energy, in contrast to the energy of residual nuclei created in the EMD process. The latter process
does not change essentially the velocity of a residual nucleus created after the emission of
few neutrons from $^{115}$In. Therefore, the removal of non-Gaussian contributions from the signals 
taken from S0 and SS together with the detection of one, two or three neutrons in the calorimeter helps
to extract signals from EMD. The fraction of events removed by such a procedure was from 2 to 3.7\%
for various sets of events.

Several measurements were performed without target. The signal in the neutron calorimeter collected 
without target, but in coincidence with signals taken from S0 and SS was considered as a background 
attributed to the interactions of beam nuclei with air, 
the walls of the vacuum tubes and other components of the setup. It was subtracted from the spectra collected in 
measurements with target in order to obtain spectra corresponding only to the interactions of beam nuclei with the target.
About 3.7\%--3.8\% of events were removed by this background subtraction.
Since the spectra obtained in separate runs with and without target corresponded to
different numbers of projectile nuclei, these spectra were appropriately weighted before the
subtraction. Namely, the spectrum obtained without target was multiplied by the ratio between
the number of beam particles which crossed the target and number of projectile nuclei in the run without target.
This ratio was calculated as 0.4--0.8, and the events leading to overflow in the last ADC channels
were discarded. The number of events with overflow collected with all five targets was from
4.8\% to 6.8\%, while it was  4\% in the run without target.

\begin{figure}[tb]
\includegraphics[width=1.0\columnwidth]{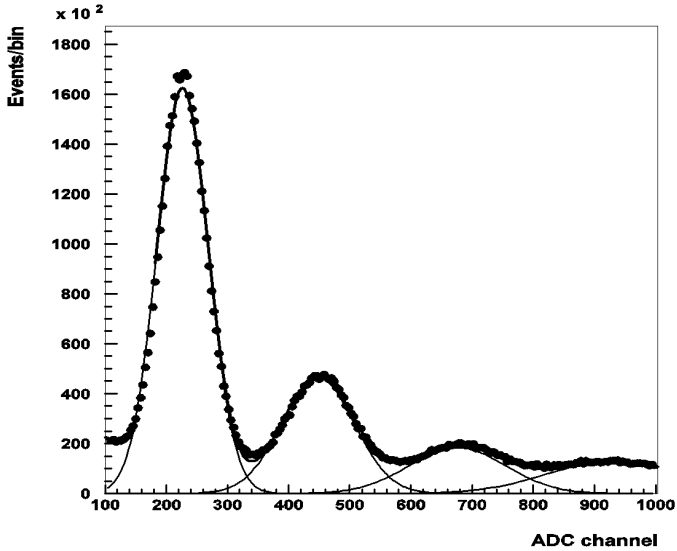}
\caption{Energy spectrum of the calorimeter for forward neutrons produced in fragmentation of
158A~GeV $^{115}$In nuclei.
}
\label{fig8:en}
\end{figure}

One of the resulting spectra obtained with $^{115}$In beam is shown in Fig.~\ref{fig8:en}.
Three distinct peaks corresponding to one, two and three neutrons which hit the calorimeter
in a single event are clearly seen. There is also a less prominent contribution from four
and more neutrons. The resulting spectra for each target were fit by a sum of Gaussians.
The average value $\mu_{1n}$ and dispersion $\sigma_{1n}$ for the 1n-peak were introduced as free
parameters of the fit. The average values for 2n, 3n and 4n events $\mu_{2n}$, $\mu_{3n}$ and $\mu_{4n}$,
and the corresponding dispersion values $\sigma_{2n}$, $\sigma_{3n}$ and $\sigma_{4n}$
were calculated through the values for 1n-peak:
$\mu_{2n}=2\mu_{1n}$, $\sigma_{2n}=\sqrt{2}\sigma_{1n}$,
$\mu_{3n}=3\mu_{1n}$, $\sigma_{3n}=\sqrt{3}\sigma_{1n}$,
$\mu_{4n}=4\mu_{1n}$, $\sigma_{4n}=2\sigma_{1n}$. 
The amplitudes of all peaks were the free parameters of the fit.

As found from the analysis of the resulting spectra, the contribution of the 3n-peak remains
essential for estimating 1n and 2n yields. In particular, the magnitude of the 2n-peak
can not be accurately estimated without accounting for the 3n-peak, as there is a large
overlap between these peaks and the corresponding Gaussians. The influence of the 4n-peak
on the 3n-peak is even more important as their widths are larger and they essentially overlap.
As seen from Fig.~\ref{fig8:en}, the 4n-peak can not be reliably identified,
but its presence is essential for extracting the 3n yield. In this work quantitative
estimates are given for 1n, 2n and 3n yields, but a small admixture of four neutrons was
also taken into account when fitting the full spectra.

The CERNLIB library was used in fitting the spectra. The area under each peak was calculated by
using respective CERNLIB functions and it was assumed proportional to the number of events of
each kind (1n, 2n or 3n). The absolute number $N$ of events with a given number of neutrons
(1, 2 or 3) produced in EMD depends on the target thickness $d$ and the mean free
pass length $\lambda$ with respect to EMD:
\begin{equation}
N=N_0(1-{\rm e}^{-d/\lambda})\ .
\label{eq:attenuation}
\end{equation}
The number of projectile nuclei is denoted here as $N_0$, $\lambda=1/n\sigma$ with
$n$ defined as the density of nuclei in the target material and $\sigma$ as the corresponding
(1n, 2n and 3n) EMD cross sections.

The number of projectiles $N_0$ was counted by S0. The cross sections extracted
from~(\ref{eq:attenuation}) are listed in Table~\ref{table3:uncorrected}. Several corrections
which take into account the absorption of beam nuclei due to hadronic fragmentation and
secondary nuclear reactions induced by emitted neutrons will be applied to these
values in the following sections.
\begin{table}[htb]
\begin{center}
\caption{Uncorrected EMD cross sections. Statistical uncertainties these values
are negligible.}
\begin{tabular}{|c|c|c|}
\hline
  Target     &        Channel     &  Cross section \\
             &                  &   (barns)      \\
\hline
     & 1n & 0.395 \\
\cline{2-3}
 Al  & 2n & 0.167  \\
\cline{2-3}
     & 3n & 0.067 \\
\hline
     & 1n & 1.541 \\
\cline{2-3}
 Cu  & 2n & 0.515 \\
\cline{2-3}
        & 3n & 0.189 \\
\hline
    & 1n & 4.090  \\
\cline{2-3}
 Sn & 2n & 1.300  \\
\cline{2-3}
    & 3n & 0.471 \\
\hline
    & 1n & 9.779 \\
\cline{2-3}
Pb thin & 2n & 3.050 \\
\cline{2-3}
    & 3n & 0.953 \\
\hline
 & 1n  & 9.750  \\
\cline{2-3}
 Pb thick & 2n  & 3.327  \\
\cline{2-3}
  & 3n  & 1.150 \\
\hline
\end{tabular}
\label{table3:uncorrected}
\end{center}
\end{table}

\section{Corrections for secondary interactions of neutrons, secondary EMD and hadronic fragmentation 
of indium nuclei}

Neutrons emitted inside the target as a result of EMD can be lost in subsequent interactions with
target material and air before they reach the neutron calorimeter. Indeed, neutrons
traverse 13~m before they reach the calorimeter and a fraction of them interact with nuclei of gases
from which air is composed. The loss of neutrons in the windows of the vacuum tubes can be
neglected, as such windows are very thin.

A projectile $^{115}$In nucleus can initiate a second EMD event following the emission of one or
two neutrons in a first EMD event. Therefore, secondary and even tertiary EMD processes are
possible inside the target. Since two subsequent emissions, each of a single neutron, are
identified by the calorimeter as a two-neutron event, the yields of 1n-events will be
artificially underestimated in favor of 2n events.

During the propagation of beam $^{115}$In nuclei and produced neutrons through the target
and air a certain sequence of phenomena which affect
the measured neutron yields takes place. According to this sequence several corrections
are applied to the raw data given in Table~\ref{table3:uncorrected}:

\begin{itemize}

\item[1.] Corrections for neutrons lost in nuclear reactions with air.

\item[2.] Corrections for neutrons lost in secondary nuclear reactions with target nuclei.
The neutrons are either stopped in the target or scattered by a large angle beyond
the acceptance of the neutron calorimeter. Due to such losses a real 2n-event is seen as
a 1n-event, and a 3n-event as a 2n- or 1n-event, respectively.
This is because of the absorption or scattering of some neutrons in multi-neutron events.

\item[3.] Corrections for secondary EMD. Two subsequent 1n-emissions from In projectile
are detected as a single 2n-event.

\item[4.] Corrections for hadronic fragmentation of $^{115}$In in the target.
This process competes with EMD and frequently leads to production of multiple nuclear fragments
without emission of forward neutrons which have the velocity close to the primary beam velocity.

\end{itemize}

Since the corrections are applicable to the measured numbers of 1n, 2n and 3n events,
they will be introduced in the next sections following the above-listed sequence.

\subsection{Corrections for neutron loss in air}\label{ninair}

The loss of neutrons in air has to be estimated first. The mean free path of $\sim158$~GeV
neutrons before they interact with air is estimated as:
\begin{equation}
\lambda_{air} = A / (N_A\times\rho_{air} \times \sigma_{nN})\ .
\label{lambdanN}
\end{equation}
In the following it is taken for simplicity that the average atomic number of air
$A=14$, which corresponds to
nitrogen, and its density $\rho_{air}=0.001205$ g/cm$^3$.  In Eq.~(\ref{lambdanN}) the Avogadro's number
$N_A=6.022\times 10^{23}$ mol$^{-1}$ and the total cross section for neutron interaction
with nitrogen $\sigma_{nN}=380$~mb~$= 0.38\times 10^{-24}$~cm$^2$ are used.
The value of $\sigma_{nN}$ was calculated for nucleon projectiles by means of the abrasion model based on the
Glauber collision theory which was successfully applied to nucleus-nucleus collisions in
Ref.~\cite{Scheidenberger}.

In the case of elastic neutron scattering on a nucleus in air it is very probable that the scattered
neutron will not hit the calorimeter. Therefore it is appropriate to use the total cross section in
calculating neutron loss in air. The cross section $\sigma_{nN}$ includes the elastic scattering cross section.
The value of $\sigma_{nN}$ can be also estimated by interpolating experimental data compiled in
Ref.~\cite{Barashenkov1993}.

Since the total cross section $\sigma_{nN}$ for nitrogen target
is not reported in the literature for 158~GeV neutrons, other combinations of the
target nucleus and neutron energy
were used in these estimations. In particular, according to the compilation~\cite{Barashenkov1993},
the total cross section for 52.7~GeV neutrons colliding with $^{16}$O amounts to  $421\pm 21$~mb~\cite{Babaev1974}
and $475\pm 44$~mb~\cite{Palevsky1967} for 1~GeV neutrons, while the total inelastic cross section for
the interaction of 100~GeV neutrons with air ($^{15}_{7.2}$Air) is 237~mb~\cite{Babayan1965}.
Since a weak energy dependence of the total cross section is expected at high neutron energies, the average of these three
measurements was calculated as $\sigma_{nN}=378$~mb. This value agrees well with $\sigma_{nN}=380$~mb calculated
by the Glauber theory, Ref.~\cite{Scheidenberger}, as explained above.

Since neutrons traverse the distance of $X=6.145$~m in air before they reach the neutron calorimeter,
$\lambda_{air}\approx 509$~m according to Eq.~(\ref{lambdanN}) and the probability
$W_{nN}$ to pass this distance without interaction is calculated as:
\begin{equation}
W_{nN}={\rm e}^{-X/\lambda_{air}}=0.988\ .
\end{equation}
The probability of the event with two neutrons escaped from the target and reached the calorimeter
without interacting in air is calculated as $W_{2n}=W_{nN}^2\approx 0.976$, while for all three
produced neutrons reached the calorimeter: $W_{3n}=W_{nN}^3\approx 0.964$.

The probability of such an event when a pair of produced neutrons leave the target
without interaction, but then one neutron interacts with air and another one reaches
the calorimeter is calculated as:
\begin{equation}
W_{1+}=2 W_{nN} (1-W_{nN})\approx 0.024\ .
\end{equation}
Due to the loss of one of two neutrons such 2n event will be detected as a 1n event.

The probability of the event with two of three neutrons interacting with air and with one
hitting the calorimeter is given by:
\begin{equation}
W_{1++} = 3 W_{nN} (1-W_{nN})^2\approx 0.0004\ .
\end{equation}
Such a 3n event is seen as a 1n event.

The probability of the event with one of three neutrons scattered in air and two others
reached the calorimeter is calculated as:
\begin{equation}
W_{2+} = 3 W_{nN}^2 (1- W_{nN}) \approx 0.035\ .
\end{equation}
Such a 3n event is seen as a 2n event.

The numbers of 1n, 2n and 3n events detected by the calorimeter and denoted as
$N_{1n-calo}$, $N_{2n-calo}$, $N_{3n-calo}$ are estimated according to the area
calculated under each of the Gaussian peaks in the calorimeter energy spectrum,
Fig.~\ref{fig8:en}. In the following the numbers of 1n, 2n and 3n events immediately
after neutron escape from the target, i.e. before neutron propagation through
air, are denoted as  $N_{1n-air}$, $N_{2n-air}$ and
$N_{3n-air}$, respectively.

It is straightforward to apply the correction for neutron absorption in air to the number of
registered 3n events: $N_{3n-air}=N_{3n-calo}/W_{3n}$.  In this calculation it is
assumed that the number of 4n events which are seen as 3n events due to the loss of one neutron
is negligible.

The number of detected neutron pairs $N_{2n-calo}$ is composed first from the number of
2n events corrected for neutron absorption, and second from the pairs of neutrons reached the
calorimeter following the loss of one neutron in 3n events:
\begin{equation}
 N_{2n-calo} = N_{2n-air} W_{2n} + N_{3n-air} W_{2+}\ .
\end{equation}
Therefore, the number of 2n events at the moment of neutron escape from the
target is calculated as:
\begin{equation}
N_{2n-air} =( N_{2n-calo} - N_{3n-air} W_{2+})/W_{2n}\ .
\end{equation}

The number of detected 1n events $N_{1n-calo}$ equals to the sum of 1n events
corrected for the neutron absorption in air, single neutrons from 2n events when one of the neutrons
is lost and single neutrons from 3n events when two other neutrons interact with air:
\begin{equation}
 N_{1n-calo}=N_{1n-air} W_{nN} + N_{2n-air} W_{1+} + N_{3n-air} W_{1++}\ .
\end{equation}
Then the number of 1n events at the moment of neutron escape from the target is calculated as:
\begin{equation}
 N_{1n-air}=(N_{1n-calo}-N_{2n-air} W_{1+} - N_{3n-air} W_{1++})/ W_{nN}\ .
\end{equation}

The equations given in this section provide $N_{*n-air}$
defined as numbers of events characterized by specific $(* = 1,2,3)$ numbers of neutrons
escaped from the target. The values of  $N_{*n-air}$ are calculated from the corresponding
counts $N_{*n-calo}$ of the neutron calorimeter. The resulting values are summarized
in Table~\ref{aircalo}.
\begin{table}[htb]
\begin{center}
\caption{Numbers of 1n, 2n and 3n events $N_{*n-calo}$, $(* = 1,2,3)$
registered by the neutron calorimeter after the propagation of neutrons in air and
the numbers of corresponding events $N_{*n-air}$ in the vicinity of the target.}
\begin{tabular}{|c|c|c|c|}
\hline
 Target   &  Channel & $N_{*n-calo}$  &  $N_{*n-air}$ \\
\hline
          &   1n   & 916500        &  918247 \\
\cline{2-4}
Al        &   2n   & 382800        &  386307 \\
\cline{2-4}
          &   3n   &   157074      &  162894 \\
\hline
          &   1n   & 1604000       &  1610282 \\
\cline{2-4}
Cu        &   2n   & 538100        &  543756 \\
\cline{2-4}
          &   3n   &   201438      & 208901  \\
\hline
          &   1n   &  634000       &   636720 \\
\cline{2-4}
Sn        &   2n   &  203000       &   205181 \\
\cline{2-4}
          &   3n   &    74738      &   77507  \\
\hline
          &   1n   & 1243000       &  1248552  \\
\cline{2-4}
Pb thin   &   2n   & 388700        &   393610 \\
\cline{2-4}
          &   3n   &   123521      &  128097 \\
\hline
          &   1n   & 1330000       &  1337861   \\
\cline{2-4}
Pb thick  &   2n   & 461600        &   466786 \\
\cline{2-4}
          &   3n   &  163885       &  169957 \\
\hline
\end{tabular}
\label{aircalo}
\end{center}
\end{table}

\subsection{Corrections for neutron loss in target material}

Once the probabilities of neutron interaction in target material are calculated,
numbers of forward neutrons produced inside the target can be estimated by correcting
the numbers of neutrons at their exit from the target, $N_{*n-air}$, obtained in Sec.~\ref{ninair}.
Similar to Eq.~(\ref{lambdanN}) written there, the mean free path
$\lambda_t$ of secondary neutrons before they interact with
target nuclei is given as:
\begin{equation}
\lambda_t = A / (N_A\times\rho\times \sigma_{ntot}) ,
\label{lambdamat}
\end{equation}
Here the density $\rho$ for each target material is used, see Table~\ref{materials}.
The total cross sections of neutron interaction with various target nuclei $\sigma_{ntot}$
were also calculated by the Glauber theory~\cite{Scheidenberger}. They are listed in
Table~\ref{xsnA} together with measured cross sections~\cite{Barashenkov1993}.
Calculated $\sigma_{ntot}$ well agrees with data for lead, but found by 15--20\% lower than the
data for other target nuclei.
\begin{table}[htb]
\begin{center}
\caption{Total cross section $\sigma_{ntot}$ (millibarns) of interaction of high-energy neutrons
with nuclei used in calculating neutron absorption in the target material. Measured cross sections
from Refs.~\cite{Babaev1974,Biel1976,Bisheva1967,Murthy1975} are given for comparison.
}
\begin{tabular}{|c|c|c|c|c|}
\hline
 Target  & Theory  &  \multicolumn{3}{|c|}{Experiment} \\
\cline{2-5}
         &    $\sigma_{ntot}$   &  Cross   &  Energy  & Expe-   \\
         &                      & section  &          & riment    \\
         &     (mb)             &   (mb)   &  (GeV)   &  \\
\hline
Al        &  510   & $623\pm 12$       & $149\pm 20$    & \cite{Biel1976}   \\
          &        & $634.8\pm 2.8$    & $179\pm 26$    &  \cite{Murthy1975}  \\
\hline
Cu        &  970   &  $1206\pm 19$   &  $149\pm 20$   & \cite{Bisheva1967}   \\
          &        &  $1223\pm 6$   &   $179\pm 26$  &  \cite{Murthy1975}   \\
\hline
Sn        &  1530   &  $1981\pm 7$   &  $54\pm 10$   &  \cite{Babaev1974}  \\
\hline
Pb        &  2960   & $3037\pm 47$    &  $149\pm 20$   &   \cite{Bisheva1967}  \\
          &        &  $2951\pm 28$    &  $179\pm 26$   &  \cite{Murthy1975}   \\
\hline
\end{tabular}
\label{xsnA}
\end{center}
\end{table}

In the following consideration it is assumed that EMD of indium nuclei
is equally probable at any depth inside the target. This means that, the points of neutron emission
are also evenly distributed along the beam path inside the target material. This assumption
is valid when the attenuation of indium beam over the target depth $d$ is relatively weak.
The probability of neutron propagation without interaction until it reaches the
depth $x$ in the target is calculated as:
\begin{equation}
 P(x)={\rm e}^{-x/\lambda_t}\ .
\end{equation}
Then the probability of neutron escape from the target without interaction is calculated as:
\begin{equation}
W_{nt}=\frac{1}{d} \int_0^d P(x){\rm d}x = \frac{1}{d}  \int_0^d {\rm e}^{-x/\lambda_t}{\rm d}x =
\frac{\lambda_t}{d} (1-{\rm e}^{-d/\lambda_t})\ .
 \end{equation}

Similar to the neutron absorption in air discussed in Sec.~\ref{ninair} the probability of escape of both
produced neutrons from the target is $W_{2nt} = W_{nt}^2$, and  $W_{3nt}=W_{nt}^3$
for all three produced neutrons. The probability of escape without interaction of
only one from a pair of produced neutrons, with the other neutron absorbed in the target
or deflected from the direction towards the calorimeter is estimated as:
\begin{equation}
 W_{1+} = 2 W_{nt} (1-W_{nt})\ .
\end{equation}

The probability of the event with two of three produced neutrons which leave the target
without interaction, while the third one is absorbed or deflected, is calculated as:
\begin{equation}
 W_{2+}= 3 W_{nt}^2 (1-W_{nt})\ .
\end{equation}

Finally, for the event with the escape of only one of the three produced
neutrons the probability is given by:
\begin{equation}
 W_{1++}=3 W_{nt} (1-W_{nt})^2\ .
\end{equation}
The above-defined probability values are summarized in Table~\ref{wnt} together with $\lambda_t$
for each target used in our measurements.
\begin{table*}[htb]
\begin{center}
\caption{The mean free path of neutrons, the probabilities of absorption for neutrons produced
in 1n, 2n and 3n events in various targets used in this experiment}
\begin{tabular}{|c|c|c|c|c|c|c|c|}
\hline
 Target  & $\lambda_t$ & $W_{nt}$ &  $W_{2nt}$ &  $W_{3nt}$ &  $W_{1+}$  &  $W_{2+}$ &  $W_{1++}$   \\
         &    (cm)     &          &            &            &            &           &              \\
\hline
  Al     &  32.560     &   0.971  &   0.943    &   0.916    &   0.057    &  0.061    &   0.0013     \\
  Cu     &  12.228     &   0.982  &   0.964    &   0.947    &   0.036    &  0.053    &   0.0001     \\
  Sn     &  17.668     &   0.990  &   0.981    &   0.970    &   0.019    &  0.028    &   0.0002     \\
Pb thin  &  10.281     &   0.993  &   0.987    &   0.979    &   0.013    &  0.019    &   0.0001     \\
Pb thick &  10.281     &   0.986  &   0.972    &   0.959    &   0.028    &  0.041    &   0.0006     \\
\hline
\end{tabular}
\label{wnt}
\end{center}
\end{table*}

The probabilities listed in Table~\ref{wnt} can be used now in calculating numbers of true
1n, 2n and 3n events inside the target, $N_{1n-target}$, $N_{2n-target}$ and $N_{3n-target}$, from
the numbers of events $N_{1n-air}$, $N_{2n-air}$ and $N_{3n-air}$ at the exit from the target.
Interactions of EMD neutrons inside the target lead to the loss of their energy and deflection
from the forward direction. As a result, interacting neutrons are not registered by the
neutron calorimeter as neutrons from EMD with their characteristic energy close to the beam
energy per nucleon.

For 3n events:
\begin{equation}
 N_{3n-air}= N_{3n-target} W_{3nt}\ ,
\end{equation}
and, respectively:
\begin{equation}
 N_{3n-target}= N_{3n-air}/W_{3nt}\ .
\end{equation}

The events with two forward neutrons leaving the target include true 2n events
corrected for neutron absorption and also 3n events with one neutron lost in interactions with
target material:
\begin{equation}
N_{2n-air}= N_{2n-target} W_{2nt} + N_{3n-target} W_{2+}\ ,
\end{equation}
and, respectively:
\begin{equation}
 N_{2n-target}=(N_{2n-air} - N_{3n-target} W_{2+})/W_{2nt}\ .
\end{equation}

The number of 1n events at the exit from the target surface consists of true 1n events with
its number corrected for absorption. In addition, it includes single neutrons from those 2n and 3n
events, where one or two neutrons, respectively, were absorbed or scattered in the target:
\begin{equation}
 N_{1n-air}= N_{1n-target} W_{nt} + N_{2n-target} W_{1+} + N_{3n-target} W_{1++}\ .
\end{equation}
This gives:
\begin{equation}
 N_{1n-target}=(N_{1n-air} -N_{2n-target} W_{1+} - N_{3n-target} W_{1++})/W_{nt}\ .
\end{equation}

The values of $N_{1n-air}$,  $N_{2n-air}$,  $N_{3n-air}$ and $N_{1n-target}$,
$N_{2n-target}$, $N_{3n-target}$ for each target are listed in Table~\ref{airtarget}.

\begin{table}[htb]
\begin{center}
\caption{Numbers of events $N_{*n-air}$ with given number of neutrons
$(* = 1, 2, 3)$ which leave the target in the direction of the neutron calorimeter and
true numbers $N_{*n-target}$ of events inside the target for each neutron multiplicity.}
\begin{tabular}{|c|c|c|c|}
\hline
Target& Channel     &   $N_{*n-air}$   &   $N_{*n-target}$  \\
\hline
     &       1n   &     918247       &    921255         \\
\cline{2-4}
 Al  &       2n   &     386307       &    392140         \\
\cline{2-4}
     &       3n   &     162894       &    173640         \\
\hline
     &       1n   &    1610282       &    1619825         \\
\cline{2-4}
 Cu  &       2n   &    543756        &    552057          \\
\cline{2-4}
     &       3n   &    208901        &    220744         \\
\hline
     &       1n   &    636720        &    638890          \\
\cline{2-4}
 Sn &        2n   &    205181        &    206875          \\
\cline{2-4}
    &        3n   &    77507         &     79775          \\
\hline
    &        1n   &    1248552       &    1251570         \\
\cline{2-4}
Pb thin&     2n   &    393610        &    396245          \\
\cline{2-4}
    &        3n   &    128097        &    130643           \\
\hline
    &        1n   &    1337861       &    1343585           \\
\cline{2-4}
 Pb thick &  2n   &    466786        &     472758          \\
\cline{2-4}
          &  3n   &    169957        &     177388           \\
\hline
\end{tabular}
\label{airtarget}
\end{center}
\end{table}

\subsection{Corrections for multiple EMD events}

Following EMD of a beam nucleus leading to the emission of few neutrons from $^{115}$In,
a further EMD process of a residual nucleus can also take place. Therefore, two subsequent EMD
events can take place in the target after the entry of a single beam nucleus, and appropriate
corrections to $N_{*n-target}$ are necessary. One can assume that the total EMD cross section for
the nuclear residue in the second EMD process is approximately equal to the corresponding cross
section for primary $^{115}$In nuclei. Due to a relatively small target thickness the third EMD process
can be neglected.

The mean free path of $^{115}$In in target material with respect to the EMD process with emission
of a single neutron, $\lambda_{1n}$, is calculated as:
\begin{equation}
 \lambda_{1n}=A/(N_A \times \rho \times \sigma_{1n})\ ,
\end{equation}
where $\sigma_{1n}$ is the cross section of 1n-emission as a result of EMD, which is calculated by the
RELDIS model, see Table~\ref{table1:theory}, and $\rho$ is
the density of target material, see Table~\ref{materials}. One can also assume that
the probabilities of single and double EMD
obey a Poisson distribution characterized by the average number of 1n events:
\begin{equation}
 \mu=1-{\rm e}^{-d/\lambda_{1n}}\ ,
\end{equation}
where $d$ is the target thickness.
Therefore, the probability of the event when a beam nucleus propagates through the target without
interaction is calculated as:
\begin{equation}
 P_0={\rm e}^{-\mu} ,
\end{equation}
while the probability of a single 1n EMD event is:
\begin{equation}
  P_1=\mu {\rm e}^{-\mu}\ ,
\end{equation}
The probability of two subsequent 1n EMD events is:
\begin{equation}
  P_2=\frac{\mu^2}{2}{\rm e}^{-\mu}\ .
\end{equation}

Following these relations the ratio between double and single EMD events with 1n emission equals to
$P_2/P_1=\mu/2$. Since a pair of two subsequent 1n emissions inside the target is identified by the calorimeter
as a single 2n event, the number of true 1n events $N_{1n}$ is calculated by introducing a correction to
$N_{1n-target}$:
\begin{equation}
 N_{1n}=N_{1n-target} (1+2P_2/P_1).
\end{equation}
Finally, the number of true 2n events $N_{2n}$ is calculated by subtracting the number of
double 1n events:
\begin{equation}
 N_{2n}=N_{2n-target} - N_{1n-target} P_2/P_1 .
\end{equation}

The results for $N_{1n}$ and $N_{2n}$ obtained after this correction
are summarized in Table~\ref{nfinal}.
\begin{table}[htb]
\begin{center}
\caption{Mean free path $\lambda_{1n}$ of 158A~GeV $^{115}$In in target material
with respect to 1n emission in EMD, the average number $\mu$ of such events for a given target, and the 
numbers, $N_{1n}$ and $N_{2n}$, of true 1n and 2n events after correction for
for multiple 1n events.}
\begin{tabular}{|c|c|c|c|c|}
\hline
Target&       Channel  &   $\lambda_{1n}$  &  $\mu$  & $N_{*n}$  \\
      &                &         (cm)      &         &           \\
\hline
 Al       &       1n   &   41.828         &  0.0329 &  982910    \\
\cline{2-5}
          &       2n   &                  &         &  361312     \\
\hline
 Cu       &       1n   &   6.217          & 0.0698  &  1854084   \\
\cline{2-5}
          &       2n   &                  &         &  434927    \\
\hline
 Sn       &        1n   &   4.94          &  0.0665 &  726747    \\
\cline{2-5}
          &        2n   &                 &         &  162947     \\
\hline
 Pb thin  &        1n   &   2.166         &  0.0604 &  1407463    \\
\cline{2-5}
          &        2n   &                 &         &  318299     \\
\hline
Pb thick  &        1n   &   2.166         &  0.1267 &  1707008     \\
\cline{2-5}
          &        2n   &                 &         &  291046      \\
\hline
\end{tabular}
\label{nfinal}
\end{center}
\end{table}

\subsection{Corrections for hadronic fragmentation of $^{115}$In in the target}\label{hadfrag}

Beam nuclei propagating through the target initiate electromagnetic as well as hadronic
collisions with target nuclei. A part of beam nuclei is destroyed in hadronic interactions.
Therefore, the beam of $^{115}$In nuclei is attenuated due to hadronic
nucleus-nucleus collisions, and the corresponding correction factor $P_{nuc}$
have to be applied to $N_{*n}$:
\begin{equation}
P_{nuc}={\rm e}^{-d/\lambda_{nuc}}\ ,
\end{equation}
\label{eq:atthadr}
where $\lambda_{nuc}=1/n\sigma_{nuc}$ with
$n$ defined as the density of nuclei in the target material and $\sigma_{nuc}$
as the total hadronic cross section listed in Table~\ref{table1:theory}.

The correction factors are listed in Table~\ref{pnuc} for each target. They are especially important
for Al and Cu targets.
\begin{table}[htb]
\begin{center}
\caption{The mean free path of $^{115}$In nuclei in the target materials with respect to hadronic
fragmentation and the corresponding attenuation factor.}
\begin{tabular}{|c|c|c|}
\hline
 Target  & $\lambda_{nuc}$ & $P_{nuc}$   \\
         &    (cm)     &             \\
\hline
  Al     &  4.97     &   0.755        \\
  Cu     &  2.79     &   0.851          \\
  Sn     &  5.26     &   0.937          \\
Pb thin  &  4.92     &   0.973     \\
Pb thick &  4.92     &   0.942      \\
\hline
\end{tabular}
\label{pnuc}
\end{center}
\end{table}

\section{Final results and discussion}

The final results of our study for the cross sections of 1n, 2n and 3n
emission in EMD along with the correction factors applied at each step of the analysis are
presented in Table~\ref{corrected}. As seen, the corrections are essential for 1n cross sections
measured for all targets. Due to relatively large  $\sigma_{nuc}$ compared to the EMD
cross sections, the corrections introduced for Al target are important for all EMD channels.
At the same time the corrections for 3n cross sections on Sn and Pb targets are small.
\begin{table*}[htb]
\begin{center}
\caption{Uncorrected 1n, 2n and 3n emission EMD cross sections for 158A~GeV $^{115}$In on
Al, Cu, Sn and Pb, correction factors for neutron absorption in air, in the target, for double EMD
and for hadronic processes which are progressively
applied in each column. Resulting cross sections are given in the last column.}
\begin{tabular}{|c|c|c|c|c|c|c|c|}
\hline
 Target & EMD  & Uncorrected    & air  & air     & air         & air        &   cross \\
        &  channel    &  cross section &      & + target & + target     & + target    & section  \\
        &             &   (barns)      &      &         & + double EMD & + double EMD&  (barns)   \\
        &             &                &      &         &              & + hadronic &             \\
\hline
               & 1n       & 0.395                  &1.00191 &1.00519   &1.072462     &1.421263        &   0.562                \\
\cline{2-8}
 Al           & 2n       & 0.167                  &1.00916   &1.02440     &0.943868     &1.250846        &   0.209                \\
\cline{2-8}
             & 3n        & 0.0667                 &1.00191   &1.00519     &      -           &1.332111        &   0.0888               \\
\hline
            & 1n         & 1.541                  &1.00392  &1.00987   &1.155918      &1.358164        &  2.093                \\
\cline{2-8}
 Cu       & 2n         & 0.515                  &1.01051  &1.02594   &0.808267      &0.949686         &  0.489               \\
\cline{2-8}
            & 3n         & 0.189                  &1.00392  &1.00987   &      -           &1.186563        &  0.224                 \\
\hline
           & 1n          & 4.090                  &1.00429  &1.00771   &1.146285     &1.222837        &  5.001                \\
\cline{2-8}
 Sn      & 2n          & 1.300                  &1.01074  &1.01909   &0.802694     &0.856300        &  1.113                \\
\cline{2-8}
          & 3n           & 0.471                 &1.00429  &1.00771   &      -           &1.075008        &  0.507                \\
\hline
         & 1n            & 9.779                  &1.00447 &1.00689   &1.132306     &1.163778        &  11.381              \\
\cline{2-8}
Pb thin & 2n         & 3.050                  &1.01263 &1.01941   &0.818879     &0.841640        &   2.567              \\
\cline{2-8}
          & 3n          & 0.953                  &1.00447 &1.00689  &     -           &1.034876        &   0.986               \\
\hline
         & 1n          & 9.750                   &1.00365 &1.00794 &1.2805753   &1.3593603        &  13.254           \\
\cline{2-8}
 Pb thick & 2n    &  3.327                   &1.01124 &1.02417   &0.630515     &0.669306       & 2.227                    \\
\cline{2-8}
             & 3n     & 1.150                    &1.00365  &1.00794   &      -           &1.069952        &  1.230       \\
\hline
\end{tabular}
\label{corrected}
\end{center}
\end{table*}

The measured cross sections are compared to ones calculated with RELDIS in
Table~\ref{reldis_comparision}. Calculated cross sections itself typically have uncertainties at
the level of 5-7\% due to uncertainties in photonuclear cross sections used as input in
calculations.

Due to a large number ($>10^4$) of events registered by the calorimeter
for each neutron multiplicity the statistical uncertainties of our measurements are small,
below 0.3\%, and can be neglected. Therefore, only systematical uncertainties are
listed in Table~\ref{reldis_comparision}. They are attributed mostly to uncertainties in
corrections applied to account for the absorption of beam nuclei and produced neutrons. Such
corrections introduced to data obtained with both thin and thick Pb targets
should give same results in principle.
Once the cross sections estimated from measurements with this pair of targets diverge,
this difference can be considered as a systematical uncertainty of our analysis.
As follows from Table~\ref{reldis_comparision}, this difference is  $\sim 15$\%
for 1n and 2n channels and $\sim 20$\% for 3n channel.  The same relative uncertainties
were attributed to the cross sections measured with other three targets.

\begin{table}[htb]
\begin{center}
\caption{Measured 1n, 2n and 3n emission EMD cross sections for for 158A~GeV $^{115}$In on
Al, Cu, Sn and Pb and cross sections calculated with RELDIS. Experimental errors are only due to
systematical uncertainties of corrections introduced for the absorption of beam nuclei and produced
neutrons in the target material and air.}
\begin{tabular}{|c|c|c|c|}
\hline
 Target &  EMD       &  Experiment  & RELDIS    \\
        &  channel      & (barns)   &  (barns)  \\
\hline
    & 1n & 0.56 $\pm$ 0.08 & 0.40 \\
\cline{2-4}
 Al & 2n &  0.21 $\pm$ 0.03 & 0.09 \\
\cline{2-4}
    & 3n & 0.09 $\pm$ 0.02 & 0.03 \\
\hline
    & 1n & 2.09 $\pm$ 0.31 & 1.91 \\
\cline{2-4}
 Cu   & 2n & 0.49 $\pm$ 0.07 & 0.45 \\
\cline{2-4}
    & 3n & 0.22 $\pm$ 0.04 & 0.13 \\
\hline
    & 1n & 5.00 $\pm$ 0.75 & 5.47 \\
\cline{2-4}
 Sn & 2n & 1.11 $\pm$ 0.17 & 1.28 \\
\cline{2-4}
    & 3n &  0.5 $\pm$0.1 & 0.38 \\
\hline
    & 1n & 11.4 $\pm$ 1.7 & 14.05 \\
\cline{2-4}
    Pb thin & 2n & 2.57 $\pm$ 0.38 & 3.32 \\
\cline{2-4}
         & 3n   & 0.99 $\pm$ 0.20 & 1.03 \\
\hline
   &   1n &  13.25 $\pm$ 2.00 & 14.05 \\
\cline{2-4}
    Pb thick & 2n  & 2.23 $\pm$ 0.33 & 3.32 \\
\cline{2-4}
            &  3n & 1.23 $\pm$ 0.25 & 1.03 \\
\hline
\end{tabular}
\label{reldis_comparision}
\end{center}
\end{table}

As seen from Table~\ref{reldis_comparision}, the measured cross sections of forward neutron emission
in collisions of $^{115}$In with Cu, Sn and Pb agree well with RELDIS taking
into account the uncertainties of measurements listed in the table.
However, the data for Al for all neutron multiplicities are noticeably larger compared to
RELDIS results for this target nucleus.

\begin{figure}[tb]
\includegraphics[width=0.95\columnwidth]{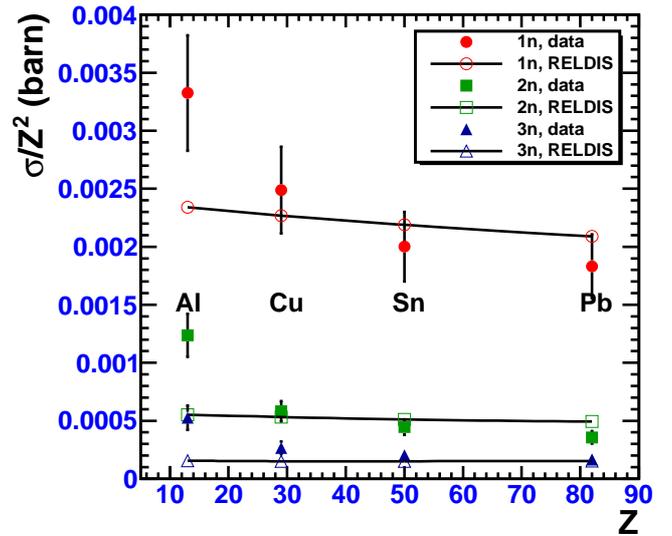}
\caption{Cross sections of emission of one, two and three neutrons by 158A~GeV $^{115}$In
as a function of the target nuclei charge $Z$. Measured 1n, 2n and 3n cross sections are shown by
circles, squares and triangles, respectively. Open symbols connected by a solid line
represent RELDIS results. All values are divided by $Z^2$ to demonstrate the characteristic dependence of
the EMD cross sections.
}
\label{fig9:SigbyZ2}
\end{figure}

A convenient presentation of the data and calculations is
given in Fig.~\ref{fig9:SigbyZ2}, where all cross sections are divided by $Z^2$, the square of
charge of target nuclei, and plotted as a function of $Z$.
As expected,  the spectrum of equivalent photons is approximately
proportional to $Z^2$ of the target nuclei (Al, Cu, Sn, or Pb), which emit photons absorbed
by $^{115}$In. This is illustrated by a flat dependence of RELDIS results divided by $Z^2$
as a function of $Z$ with its slight increase for lighter targets. This increase is
explained by a wider range of impact parameters available in ultraperipheral collisions with
light targets defined by $b>R_1+R_2$, see Sec.~\ref{photonuclear}.  This provides higher $E_{max}$
for Al and Cu compared to Pb target. Despite of this peculiarity a flat dependence of
$\sigma/Z^2$ serves as a clear indication of the electromagnetic nature of
interactions. In particular, a similar flat dependence was observed in our studies of the
emission of forward neutrons by 30A~GeV Pb nuclei in collisions with Al, Cu, Sn and
Pb~\cite{Golubeva}.

Taking into account experimental uncertainties of the present data,
$\sigma/Z^2$ for Cu, Sn and Pb targets are consistent in general with a
flat dependence on $Z$. However, the cross sections measured with Al target are higher
than calculated with RELDIS, and also a general descending trend of $\sigma/Z^2$ is obvious
in Fig.~\ref{fig9:SigbyZ2}. As follows from
Table~\ref{table1:theory}, the cross sections of 1n, 2n and 3n emission in hadronic
interactions of $^{115}$In with Al are
comparable to the corresponding EMD cross sections. Therefore, a noticeable contribution of
hadronic events to the emission of forward neutrons by $^{115}$In in collisions with Al is expected.
Apart of the EMD process, one, two and three neutrons are also produced in peripheral grazing hadronic
nucleus-nucleus collisions. For a given projectile nucleus the cross section of grazing
collisions is proportional to the area of a thin surface rim of the target nucleus:
$\sigma \sim Z^{2/3}$. This means that $\sigma/Z^2\sim 1/Z^{4/3}$, and the descending trend
which is seen in Fig.~\ref{fig9:SigbyZ2} due the presence of hadronic events is thus explained.

\section{Conclusions}

The cross sections of emission of one, two and three forward neutrons in collisions
of 158A~GeV $^{115}$In projectiles with Al, Cu, Sn and Pb target nuclei are
measured. The collected data are compared with the cross sections of neutron emission in
electromagnetic dissociation of $^{115}$In calculated with the RELDIS model.
The measured cross sections agree with the cross sections calculated by RELDIS
for ultraperipheral collisions with Cu, Sn and Pb targets within estimated
uncertainties of the measurements. Such systematic uncertainties stem from the corrections to raw
data accounting for interactions of projectile nuclei and neutrons with target
material and air. The validity of the RELDIS model for simulating electromagnetic
dissociation of ultrarelativistic  $^{115}$In nuclei is demonstrated.

The measured cross sections are
attributed mainly to the electromagnetic dissociation process, as
their values divided by $Z^2$ of the target nucleus demonstrate a weak dependence on $Z$.
The excess of neutron emission cross section measured in In-Al collisions with respect to
calculated EMD cross section indicates a noticeable contribution for grazing In-Al collisions.
This is in contrast to our previous experiment~\cite{Golubeva} conducted
with 30A~GeV $^{208}$Pb beam colliding with the same targets where the contribution of
grazing collisions was found relatively small with respect to EMD also for Al target.

As demonstrated recently~\cite{Oppedisano:2011rx,AbelevALICE2012}, ultraperipheral collisions of lead
nuclei followed by emission of forward neutrons play a certain role at the LHC collider due to their
large cross sections. Forward neutron emission can be used to monitor the LHC luminosity,
providing that such cross sections are known with sufficient accuracy, e.g.,
calculated with RELDIS.

As shown~\cite{Oppedisano:2011rx,AbelevALICE2012}, such cross sections are indeed
reliably predicted by RELDIS, once this model has been previously validated
by comparison with the data~\cite{Golubeva} also collected with ultrarelativistic Pb projectiles, but
at lower energies.

Our present results confirm the accuracy of RELDIS in describing the cross sections of forward
neutron emission by medium-weight ultrarelativistic $^{115}$In nuclei.
Therefore, the RELDIS model can be also used to plan future experiments at the LHC, possibly
with beams of medium-weigh nuclei, like  $^{115}$In. New experiments on photoabsorption on various
target nuclei are very desirable, as they help to improve the accuracy of nuclear data used to model
EMD.

This work was supported by the Russian Foundation for Basic Research,
grant No. 12-02-91508-CERN and by the Program of the Russian Academy of
Sciences RAS-CERN. The authors are very grateful to V.V.~Varlamov for discussions of photonuclear
data on $^{115}$In.

\end{document}